# JFS-CryoMem: A Cryogenic Memory with Voltage-Controlled Superconducting Devices and Femtojoule-Scale Write/Read Energies

Md Rahatul Islam Udoy[1†], Md Mazharul Islam[1†], Juan P. Mendez[2], Denis Mamaluy[2], Aaron J. Muhowski[2], Samuel Hawkins[2], William M. Martinez[2], Courtney Sovinec[2], Wei Pan[3], and Ahmedullah Aziz[1*]
[1]Department of Electrical Engineering and Computer Science, University of Tennessee, Knoxville, TN 37996, USA
[2]Sandia National Laboratories, Albuquerque, NM 87123, USA
[3]Sandia National Laboratories, Livermore, CA 94550, USA
[†]Equal Contribution, *Corresponding Author's Email: aziz@utk.edu.

**Scalable cryogenic systems require memory that combines nonvolatile storage, selective access, low thermal disturbance, and compatibility with superconducting electronics. We present a cryogenic memory architecture that integrates a voltage-controlled Josephson junction field-effect transistor (JJFET) selector with a ferroelectric superconducting quantum interference device (FeSQUID) storage element, hereafter termed JFS-CryoMem. The JJFET provides gate-controlled cell selection, whereas the FeSQUID stores information in stable remanent-polarization states. JFS-CryoMem features separate read and write path mechanisms that support nondestructive readout and independent optimization of programming and sensing conditions. The architecture is evaluated using experimentally calibrated compact models that reproduce the measured electrical characteristics of both constituent devices. We demonstrate selective programming using a half-bias scheme, nonvolatile state retention, and distinguishable readout in a 4 × 4 array while accounting for the selected cell and all unselected parallel branches. We then extend the analysis to arrays up to 16 × 16 and examine how array scaling alters current distribution, column-equivalent resistance, readout separation, required bitline current, and read energy. The results reveal the principal sensing and energy tradeoffs associated with larger arrays and identify the operating conditions required to preserve read distinguishability as the array grows. JFS-CryoMem provides a device-to-array framework for cryogenic memory in quantum, high-performance, and space-oriented computing systems.**

Cryogenic electronics are emerging as an enabling platform for quantum computing, high-performance computing, and space applications[1,2]. In current quantum-computing systems, superconducting qubits operate at millikelvin temperatures, whereas much of the associated control and memory hardware remains at room temperature[3–5]. Scaling such an architecture toward thousands of qubits would require a large number of long interconnects extending from the 300 K stage to the millikelvin stage, resulting in substantial wiring complexity, communication latency, and heat leakage across a temperature difference approaching 300 K[6,7]. Integrating the control processor and memory at an intermediate cryogenic stage, such as 4.2 K, brings these components physically and thermally closer to the qubits[1,8]. This arrangement enables shorter and higher-density low-loss interconnects while reducing the thermal gradient between the supporting electronics and the qubits from approximately 300 K to only a few kelvin[9,10]. Cryogenic processors and memories can therefore alleviate the interconnect and thermal bottlenecks that currently limit the scalability of quantum-computing systems[11,12].

For high-performance computing, superconducting circuits are attractive because their ultrafast and energy-efficient switching characteristics offer a potential route beyond the power and performance limitations of conventional CMOS[13–15].

Cryogenic electronics are also relevant to deep-space and planetary missions, where electronic systems may be exposed to naturally low-temperature environments[16,17]. Devices capable of operating directly under such conditions can reduce the need for heaters and their associated power, mass, and supporting structures. These applications collectively motivate efficient processors and nonvolatile memories that can operate reliably within cryogenic environments.

Nonvolatile memory is particularly attractive because it retains information after the programming bias is removed, eliminating continuous refresh and reducing standby power[18–20]. Recent cryogenic-memory research has explored semiconductor, magnetic, and fully superconducting approaches[1]. Cryo-CMOS SRAM and DRAM have demonstrated operation near 4 K[21], while hybrid memories have combined magnetic tunnel junctions with superconducting heater cryotron selectors[22]. Fully superconducting nanowire memories have also achieved array-level operation and high integration density[23].

Despite extensive progress, superconducting, nonsuperconducting, and hybrid memories intended for operation at or below 4 K continue to face challenges including limited scalability, process complexity, bulky peripherals, array-level interference, volatility, and speed incompatibility[1,2,24]. For example, many existing designs still involve thermal activation, destructive readout, interface complexity, or limited scalability. In our previous work, we used a heater cryotron (hTron) as a selector device in the memory array[25]. The hTron offers useful characteristics. However, its switching mechanism relies on localized heating. Although effective for cell selection, intentional heat generation introduces an additional thermal consideration where cooling capacity is limited[26–28]. This motivates us to design a cryogenic memory using a heater-free, voltage-controlled selector.

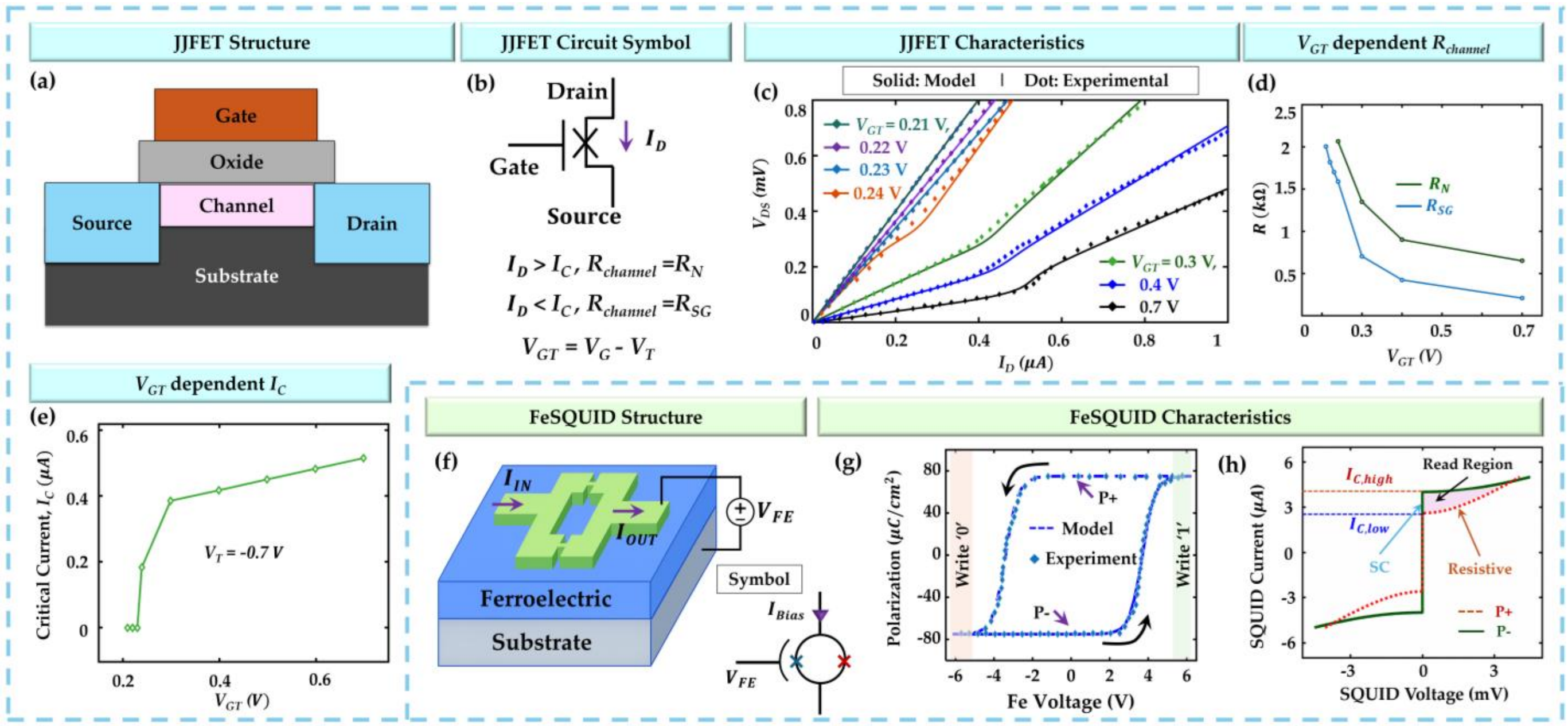

**Fig. 1. Constituent devices and experimentally calibrated characteristics of the JFS-CryoMem cell.** (a) Schematic cross section and (b) circuit representation of the voltage-controlled Josephson junction field-effect transistor (JJFET) selector. The channel exhibits the subgap resistance $R_{SG}$ when $|I_D| < I_C$ and the normal-state resistance $R_N$ when $|I_D| \geq I_C$. (c) Comparison between the modeled and experimental $I_D - V_{DS}$ characteristics for different $V_{GT}$ values. (d) Dependence of $R_{SG}$ and $R_N$ on gate overdrive voltage and (e) modulation of the critical current $I_C$ by $V_{GT}$. (f) Schematic structure and circuit symbol of the ferroelectric superconducting quantum interference device (FeSQUID), showing the ferroelectric programming voltage $V_{FE}$ and the SQUID bias path. (g) Modeled and experimental polarization–voltage hysteresis characteristics, illustrating the P− and P+ remanent-polarization states corresponding to write '0' and write '1', respectively. (h) Polarization-dependent SQUID current–voltage characteristics. The two stored states exhibit distinct critical currents, $I_{C,low}$ and $I_{C,high}$, defining a read-current range in which one state remains superconducting while the other enters the resistive regime.

Here, we propose a Josephson junction field-effect transistor (JJFET) and ferroelectric superconducting quantum interference device (FeSQUID) based cryogenic memory (JFS-CryoMem), which is heater-free and nonvolatile. We choose JJFET as the selector because it provides gate-voltage control over the critical current of a superconducting weak-link channel[29,30]. In particular, the quantum-enhanced InAs/GaSb JJFET exhibits a reported gain factor of approximately 0.06, more than 50 times that of conventional InAs-based JJFETs, providing substantially stronger gate modulation[29,31]. The FeSQUID is well suited as the storage device because its write and read operations occur through distinct physical paths, thus achieving the separate read write path (SRWP) advantages[32]. This allows programming and sensing conditions to be optimized independently and reduces read disturbance[1]. Our analysis is experimentally grounded: the JJFET compact model is calibrated to measured electrical characteristics[29], and the FeSQUID semi-physical model reproduces experimentally observed ferroelectric and superconducting behavior[33]. We investigate the complete array operation up to 16×16 array. We further examine how scaling affects current distribution, readout distinguishability, and power requirements.

## JJFET and FeSQUID

The proposed cryogenic memory cell integrates a Josephson junction field-effect transistor (JJFET) as the selector device and a ferroelectric superconducting quantum interference device (FeSQUID) as the nonvolatile storage element. Figures 1(a)-1(e) summarize the structure, electrical characteristics, and compact modeling of the JJFET, while Figs. 1(f)-1(h) present the corresponding characteristics of the FeSQUID.

The quantum-enhanced JJFET consists of a zero-gap InAs/GaSb heterostructure channel contacted by superconducting tantalum electrodes, with an effective channel length of approximately 500 nm[29]. The gate voltage modulates the carrier density and superconducting coherence within the channel, thereby controlling its critical current, $I_C$. As shown in Fig. 1(e), a sharp onset of $I_C$ occurs near a gate-overdrive voltage $V_{GT} = V_G - V_T = 0.24\ V$, where $V_T = -0.7\ V$. Below this voltage $I_C$ approaches zero and the device remains in a high-resistance state. Increasing $V_{GT}$ produces a rapid increase in $I_C$, providing the gate-controlled conduction required for cell selection.

The JJFET exhibits two distinct conduction regimes depending on the drain current. For $|I_D| < I_C$, the device operates in a low-resistance subgap regime characterized by $R_{SG}$. When $|I_D| \geq I_C$, the channel switches to the normal-state resistance $R_N$. Both resistances vary with $V_{GT}$, as shown in Fig. 1(d), and the drain voltage is expressed as

$$V_{DS} = \begin{cases} I_D R_{SG}(V_{GT}), & |I_D| < I_C(V_{GT}), \\ I_D R_N(V_{GT}), & |I_D| \geq I_C(V_{GT}). \end{cases}$$

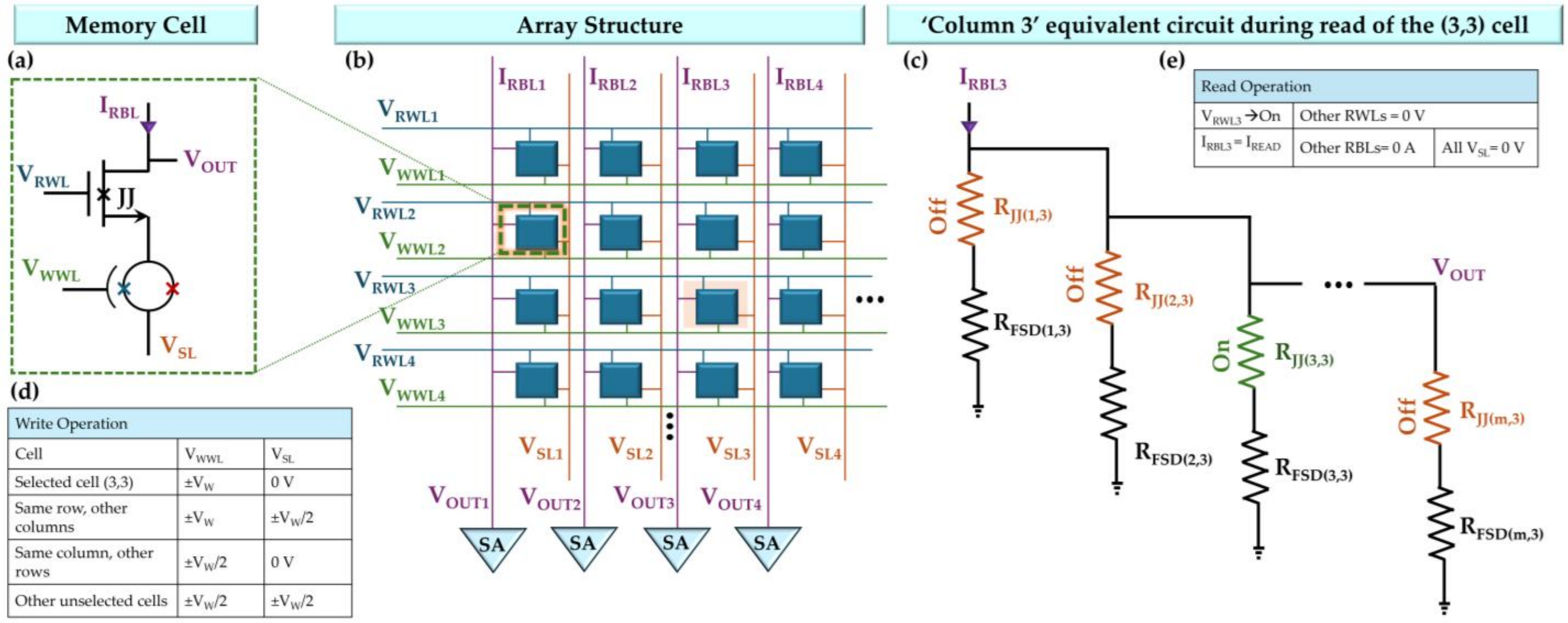

**Fig. 2. Cell organization and array-level biasing of the JFS-CryoMem architecture.** (a) Memory cell comprising a JJFET selector and a FeSQUID storage element. RWL, RBL, WWL, and SL denote the read word line, read bit line, write word line, and source line, respectively, while $V_{OUT}$ is the column output voltage. (b) Array structure with shared row and column interconnects and a sense amplifier (SA) connected to each column output. The highlighted cell indicates the selected (3,3) cell. (c) Equivalent circuit of column 3 during readout of the selected cell, where $R_{JJ(i,3)}$ and $R_{FSD(i,3)}$ represent the JJFET and FeSQUID resistances, respectively, in the $i$th branch of the column. (d) Bias conditions used for the V/2 write scheme and (e) bias conditions used for the read operation.

Owing to the relatively long channel, the subgap regime retains a finite resistance rather than exhibiting an ideal zero-voltage supercurrent branch. For circuit-level analysis, we use our experimentally calibrated Verilog-A compact model, in which $R_{SG}$ and $R_N$ are represented using gate-dependent lookup tables and $I_C$ is captured using a piecewise-linear approximation. The agreement between the modeled and experimental $I_D - V_{DS}$ characteristics is shown in Fig. 1(c).

The FeSQUID consists of a planar $Mo_{80}Si_{20}$ SQUID integrated with a PZT ferroelectric layer, as illustrated in Fig. 1(f)[33]. The ferroelectric supports two stable remanent-polarization states, '$P+$' and '$P-$', which are switched by applying positive or negative voltage across the ferroelectric. The polarization-dependent interfacial charge modulates the superconducting properties of the SQUID, including its critical temperature, superconducting energy gap, and critical current. Consequently, $P+$ and $P-$ produce two distinct critical currents, denoted by $I_{C,low}$ and $I_{C,high}$, respectively. The FeSQUID state is read by applying a current satisfying $I_{C,low} < I_{READ} < I_{C,high}$. Under this condition, the $P-$ state remains superconducting and produces a near-zero terminal voltage, whereas the $P+$ state switches to the resistive regime and generates a finite voltage, as illustrated in Fig. 1(h). These two responses represent logic '0' and logic '1', respectively. Because the polarization is programmed through the ferroelectric terminal while the stored state is sensed through the SQUID conduction path, the device provides separate write and read mechanisms.

We use our previously developed physics-informed Verilog-A model for the FeSQUID[25]. The model combines a Preisach hysteresis formulation, which captures the history-dependent ferroelectric polarization switching[34], with a resistively and capacitively shunted junction model for the SQUID. The two components are self-consistently coupled through the polarization-dependent superconducting parameters. As shown in Figs. 1(g) and 1(h), the model reproduces both the ferroelectric hysteresis and the state-dependent SQUID characteristics. The resulting combination of voltage-controlled JJFET selection and nonvolatile FeSQUID storage forms the basis of the proposed JFS-CryoMem array. In this work, we consider 1 K as the operating temperature of the JFS-CryoMem architecture. This temperature is selected because it lies below the experimentally observed superconducting transition of approximately 1.2 K in the JJFET[29,30], while also remaining well below the reported critical temperatures of the FeSQUID[33]. The FeSQUID characteristics used in the compact model were experimentally measured at 3 K and are employed here as a first-order approximation at 1 K. Although the FeSQUID is expected to remain in the superconducting regime at 1 K, superconducting parameters, particularly the critical current, are temperature dependent and may therefore differ quantitatively from their measured values at 3 K. Accordingly, a temperature-dependent calibration of the FeSQUID model is left for future work.

**JJFET and FeSQUID-based Cryogenic Memory (JFS-CryoMem)**

Figure 2(a) shows the proposed JFS-CryoMem cell, which integrates a JJFET selector with an FeSQUID storage device. The read word line, RWL, controls the JJFET, while the read bit line, RBL, supplies the read current. The write word line, WWL, and source line, SL, apply the programming voltage across the ferroelectric layer of the FeSQUID. Figure 2(b) shows an $m \times n$ JFS-CryoMem array, where the cells in each column share an RBL and a common column output, $V_{OUT}$, for sensing the stored state.

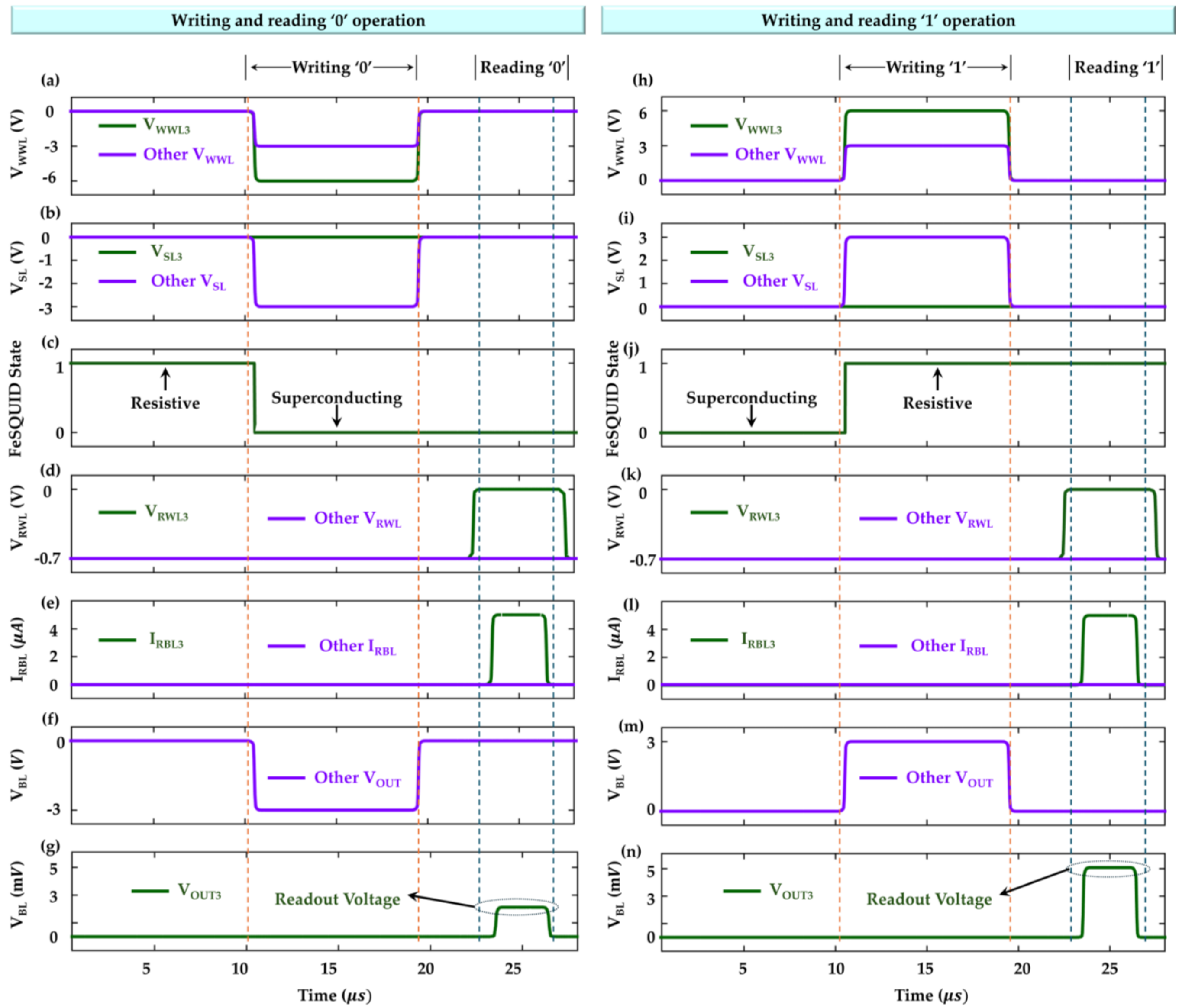

**Fig. 3. Simulated write and read operation of the selected (3,3) cell in a 4 × 4 JFS-CryoMem array.** Panels (a)-(g) show the applied bias signals, FeSQUID state transition, and column output voltage during writing and subsequent reading of logic '0', while panels (h)-(n) show the corresponding operation for logic '1'. The read operation produces distinct $V_{OUT3}$ levels for the two stored logic states. The write operation uses the $V/2$ biasing scheme so that only the selected cell receives the full programming voltage, while half-selected cells receive half of the programming voltage. After the write voltage is removed, the FeSQUID retains its programmed polarization state, demonstrating nonvolatile storage. During readout, the JJFET of the selected cell is enabled and the two stored states produce distinct column output-voltage levels, ($V_{OUT}$), without altering the stored state.

We program the array using the $V/2$ biasing scheme summarized in Fig. 2(d). To write the (3,3) cell, we apply $\pm V_W$ to $V_{WWL3}$ while grounding $V_{SL3}$. We bias the other WWLs and SLs at $\pm V_W/2$. Therefore, the selected cell receives the full programming voltage, the cells sharing either its row or column receive only half of the programming voltage, and the remaining cells receive zero voltage. Accordingly, the selected, half-selected, and unselected cells experience $\pm V_W, \pm V_W/2$, and 0 V, respectively. We choose the programming conditions such that only $\pm V_W$ exceeds the coercive voltage of the ferroelectric. Thus, only the selected cell switches its polarization. A negative programming voltage sets the selected cell to P−, representing logic '0', whereas a positive programming voltage sets it to P+, representing logic '1'.

During read operation, we apply $I_{READ}$ only to the selected column. We turn on the JJFET of the selected cell by setting $V_{RWL3}$ to 0 V, while maintaining the unselected RWLs at −0.7 V to keep their JJFETs off. Figure 2(c) shows the equivalent circuit of column 3 while reading the (3,3) cell. Each branch contains the series combination of the JJFET resistance, $R_{JJ(i,3)}$, and the FeSQUID resistance, $R_{FSD(i,3)}$. Defining the total resistance of the $i$th branch as $R_{i,3} = R_{JJ(i,3)} + R_{FSD(i,3)}$, the output voltage of column 3 is

$$V_{\mathrm{OUT3}} = I_{\mathrm{RBL3}} \left[ \sum_{i=1}^{m} \frac{1}{R_{\mathrm{JJ}(i,3)} + R_{\mathrm{FSD}(i,3)}} \right]^{-1} \tag{1}$$

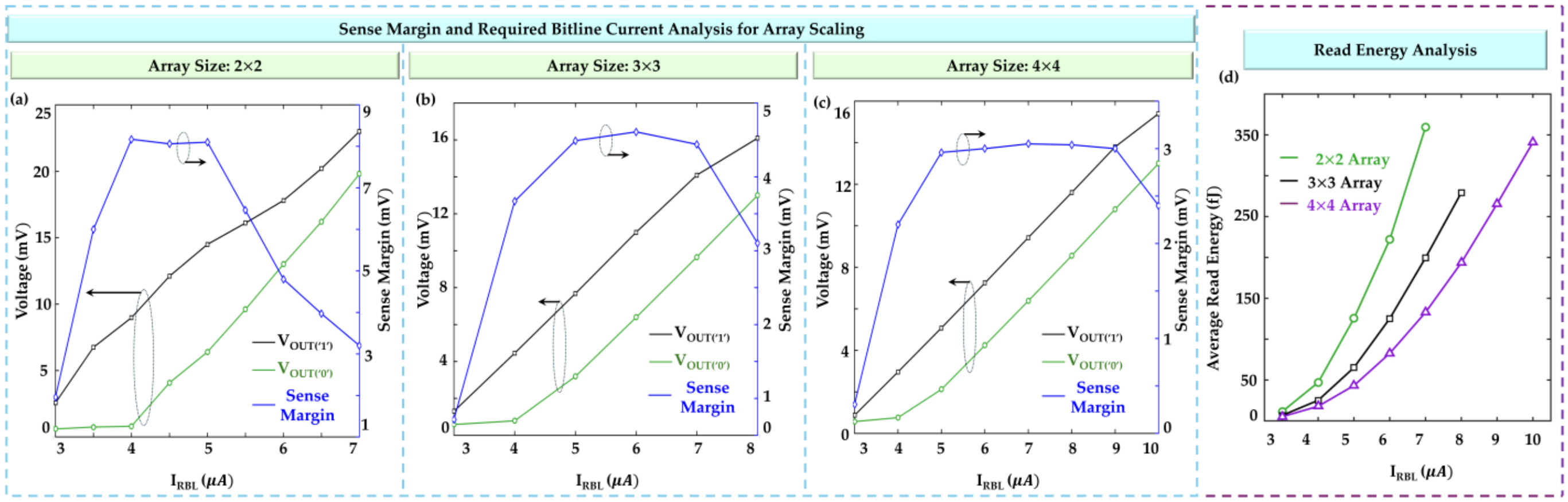

**Fig. 4. Array-scaling analysis of the JFS-CryoMem read operation**. The column output voltages corresponding to logic '0' and logic '1', together with the voltage sense margin (SM), as functions of the applied read bitline current for (a) 2 × 2, (b) 3 × 3, and (c) 4 × 4 arrays. As the array size increases, additional unselected branches divert a larger portion of the injected current, requiring a higher $I_{RBL}$ to drive the selected FeSQUID into the read region. The additional parallel branches also reduce the column-equivalent resistance and, consequently, the absolute output voltage at a given $I_{RBL}$. (d) Average read energy per cell obtained from circuit simulations as a function of $I_{RBL}$. At a fixed $I_{RBL}$, larger arrays exhibit lower read energy because of their reduced column-equivalent resistance and output voltage; however, this fixed-current comparison does not represent an equivalent sensing condition across different array sizes.

The current through the selected ON branch is therefore

$$I_{\mathrm{ON}(3,3)} = I_{\mathrm{RBL3}} \frac{\frac{1}{R_{\mathrm{JJ}(3,3)} + R_{\mathrm{FSD}(3,3)}}}{\sum_{i=1}^{m} \frac{1}{R_{\mathrm{JJ}(i,3)} + R_{\mathrm{FSD}(i,3)}}} \tag{2}$$

Because the OFF-state JJFET resistances are significantly larger than the resistance of the selected branch, the output voltage can be approximated as

$$V_{\mathrm{OUT3}} \approx I_{\mathrm{READ}} \left[ R_{\mathrm{JJ}(3,3)} + R_{\mathrm{FSD}(3,3)} \right] \tag{3}$$

The FeSQUID resistance depends on the stored polarization. For logic '0', the FeSQUID remains superconducting during read and produces the lower $V_{OUT}$ level. For logic '1', it switches to the resistive state and produces the higher $V_{OUT}$ level. The finite resistance of the ON-state JJFET results in a nonzero output even when the FeSQUID remains superconducting.

Figure 3 shows the simulated write and read operations of the selected (3,3) cell for a 4×4 memory array. During write '0', we apply −6 V to $V_{WWL3}$ and −3 V to the other WWLs and unselected SLs, while keeping $V_{SL3}$ at 0 V. The selected FeSQUID consequently switches from the resistive $P+$ state to the superconducting $P-$ state. During write '1', we reverse the bias polarities and apply +6 V and +3 V, switching the selected cell from $P-$ to $P+$. After removing the programming voltage, the FeSQUID retains its state, confirming nonvolatile storage. The subsequent read pulses produce two distinct $V_{OUT3}$ levels without altering the stored state.

The JFS-CryoMem provides separate read and write paths because the write voltage acts across the ferroelectric terminal, whereas the read current flows through the JJFET and SQUID conduction path. This separation also allows us to optimize the programming and sensing conditions independently.

Although the programming voltage reaches ±6 V while the read output remains in the millivolt range, the JFS-CryoMem does not require voltage-level matching between its write input and read output. The read voltage does not drive or program another memory cell. Therefore, the difference between the programming and read voltage ranges does not limit array operation. This differs from a logic circuit, where an output must directly drive the input of a subsequent gate. The column output of the memory array only needs to be detected by a sense amplifier, which distinguishes the two readout voltage levels and resolves the stored logic state. The design of a sense amplifier is beyond the scope of this work, however, a JJFET-based cryogenic sense amplifier has been reported in a previous work[35].

We emphasize that the millivolt-level readout remains distinguishable at the 1 K operating temperature. The thermal voltage at 1 K is $kT/q \approx 0.086$ mV. In our case, the sense margin (defined by the absolute difference between the readout voltages corresponding to logic '1' and logic '0') is ~3 mV (Fig. 3(g), (n)), which is ~35 × larger than $kT/q$. The read window is therefore well above the thermal-voltage scale. The estimated write energy is 27 fJ per cell, while the read-energy, depending on memory cell array size, is about 10 fJ. Detailed analysis is presented later in this paper.

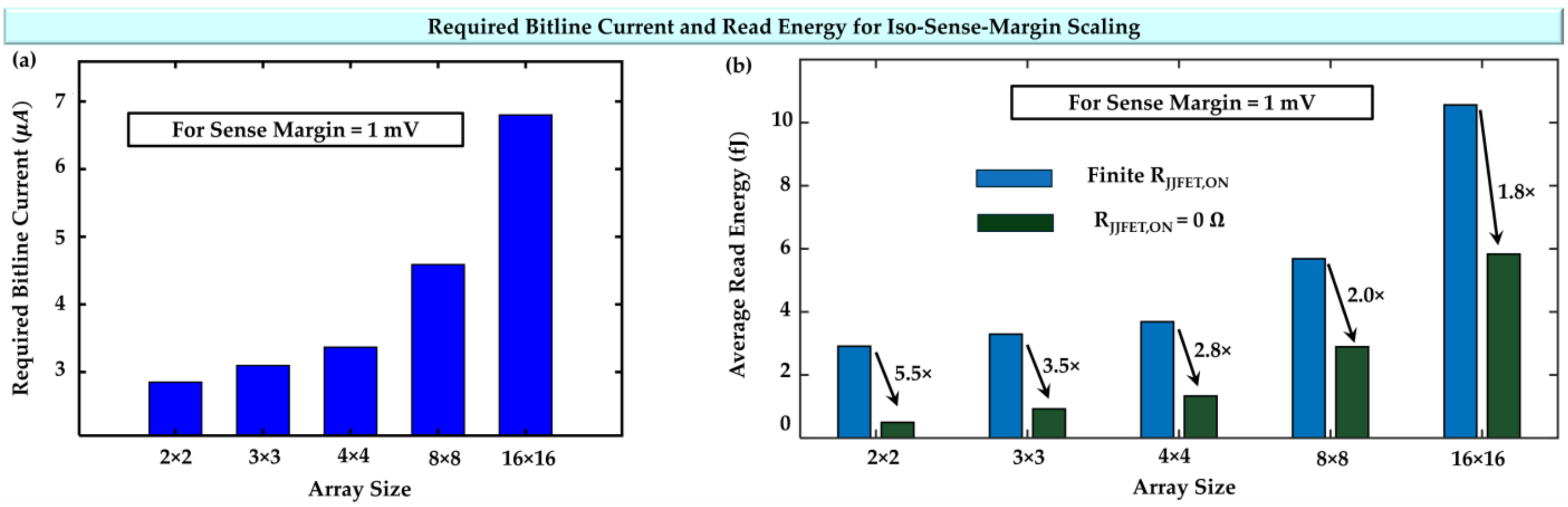

**Fig. 5. Iso-sense-margin scaling of the JFS-CryoMem array.** (a) Required read bitline current ($I_{RBL}$), for array sizes from 2 × 2 to 16 × 16 while maintaining a fixed voltage sense margin of 1 mV. As the array size increases, additional unselected branches divert a larger fraction of the applied bitline current, requiring a higher ($I_{RBL}$) to preserve the same sensing condition. (b) Corresponding average read energy obtained by integrating the simulated read power over a 2.4 μs read interval. Results are shown for the experimentally calibrated JJFET with finite ON-state resistance and for an idealized JJFET with 0 Ω ON-state resistance.

## Scaling of the JFS-CryoMem

Array scaling is important because a memory concept that operates correctly at the single-cell or small-array level must also preserve reliable readout as the number of connected cells increases. In JFS-CryoMem, enlarging the array introduces additional unselected branches that alter current distribution and column-equivalent resistance, which can reduce the current delivered to the selected cell and degrade the separation between the two read states. Therefore, scaling analysis is necessary to determine whether reliable state discrimination can be maintained in larger arrays and to quantify the additional read-current and energy requirements associated with doing so. From a broader memory-architecture perspective, larger arrays are also desirable because row/column drivers, sensing circuitry, and other peripheral components can be shared across a greater number of cells, reducing the peripheral overhead per stored bit and improving effective integration density. In this section, we evaluate how array scaling affects the sense margin, required bitline current, and read energy, as these parameters determine the practical scalability and readability of the JFS-CryoMem array. Here, the voltage sense margin is defined as the absolute difference between the readout voltages corresponding to logic '1' and logic '0'. Figures 4(a)-4(c) show the readout voltages corresponding to logic '0' and logic '1', together with the resulting voltage sense margin (SM) for the 2 × 2, 3 × 3, and 4 × 4 JFS-CryoMem arrays. We performed a complete array-level simulation for every applied $I_{RBL}$ value; therefore, each point along the x-axis represents an independent simulation of the full array, including the selected cell and all unselected parallel branches. From each simulation, we extracted the two logic-state readout voltages and calculated SM from their absolute difference. For each array size, SM initially increases with $I_{RBL}$, reaches a maximum, and then gradually decreases. This nonmonotonic behavior originates from the FeSQUID characteristics shown in Fig. 1(h). When the current through the selected cell exceeds $I_{C,low}$, the low-critical-current state enters the resistive region, while the high-critical-current state remains superconducting. The separation between the two readout voltages therefore increases. As the current increases further, this separation reaches its maximum within the read region. Once the selected-cell current approaches or exceeds $I_{C,high}$, both states increasingly exhibit resistive behavior, causing their readout voltages to converge and reducing SM.

The bitline current required to access this read region increases with array size. As described by Eq. (2), the injected current divides among all the cell branches connected to the selected column. The JJFETs in the unselected branches remain OFF and present large resistances; however, these resistances are finite, and the corresponding branches still draw a small portion of the injected current. Adding more rows therefore increases the total current diverted through the unselected branches and reduces the fraction of $I_{RBL}$ that flows through the selected ON branch. Consequently, we must apply a larger bitline current in a larger array to deliver the required current through the selected FeSQUID.

Array scaling also reduces the absolute readout voltages. Each additional row introduces another parallel branch in the selected column. Although the added branch has a high resistance because its JJFET remains OFF, it still lowers the equivalent resistance of the column. According to Eq. (1), $V_{OUT}$ is determined by the product of $I_{RBL}$ and the column-equivalent resistance. Therefore, for the same injected current, the reduction in equivalent resistance produces a smaller voltage drop as the array size increases. This trend appears in Figs. 4(a)-4(c).

Figure 4(d) shows the average read energy estimated from circuit simulations. At a fixed bitline current, the larger arrays exhibit lower read energy because their reduced column-equivalent resistance results in a smaller output voltage. However, comparing different array sizes at the same $I_{RBL}$ does not represent an equivalent sensing condition. A larger array requires a higher $I_{RBL}$ to compensate for the additional current division and maintain the same sense margin.

We therefore evaluate array scaling under an iso-sense-margin condition in Fig. 5 (up to 16×16 array). As shown in Fig. 5(a), maintaining a fixed SM of 1 mV ($\sim 3kT/q$) requires a progressively larger bitline current as the array size increases. Figure 5(b) shows the corresponding average read energy per cell. The read energy is calculated by integrating the simulated read power over a 2.4 µs read interval. Because the experimentally calibrated JJFET exhibits a finite ON-state resistance, it introduces additional dissipation during readout. To highlight this effect, we also consider an ideal case with 0 Ω on resistance of the JJFET, which reduces the read energy, indicated by the dark green bars in Fig. 5(b). We further point out here that although a larger array may consume less read energy at a fixed bitline current, maintaining the same sensing capability requires a progressively larger read current and, consequently, higher read energy. These results reveal the scaling tradeoff among array size, bitline current, sense margin, and read energy in the proposed JFS-CryoMem architecture.

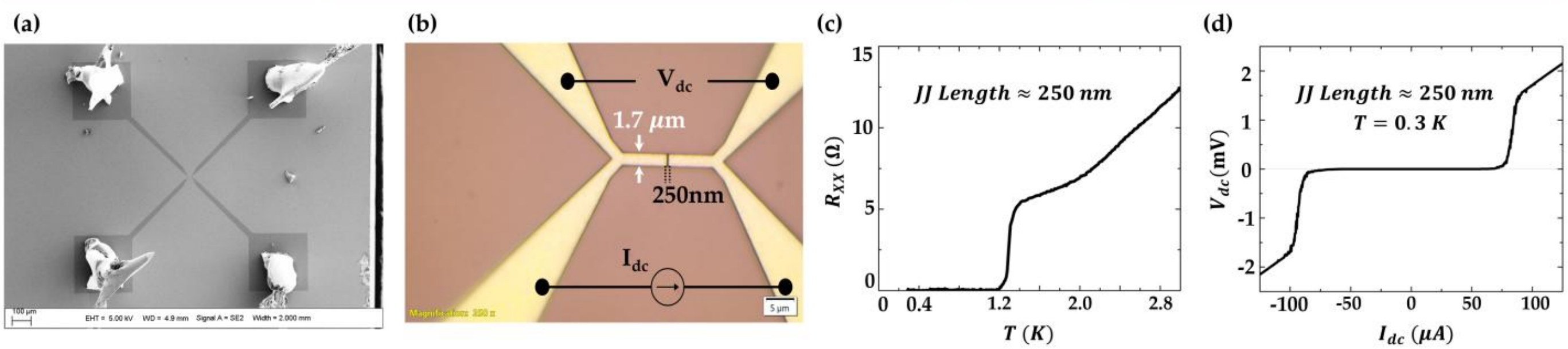

**Fig. 6. Experimental characterization of the optimized Josephson junction.** (a) Scanning electron microscopy (SEM) image of the fabricated Josephson junction (JJ) device with indium contacts. (b) Magnified optical image of the junction region, showing a junction length of approximately 250 nm and a width of approximately 1.7 µm, and schematic measurement setup for current ($I_{dc}$) – voltage ($V_{dc}$) characteristics. (c) Temperature dependence of the junction resistance, ($R_{XX}$). (d) $Idc - V_{dc}$ characteristics of the junction measured at ($T$ = 0.3) K.

## Experimental Demonstration of Zero On-Resistance

The ideal zero-on-resistance case considered in Fig. 5(b) can be approached experimentally through optimization of the Josephson junction structure. In an optimized Josephson junction (JJ) fabricated with epitaxial aluminum (Al)[30], zero on-resistance can be achieved. In our previous experiment in a JJFET, whose data is shown in Fig. 1, the length of JJ was approximately 500 nm, and tantalum (Ta) was used as the superconducting electrodes[29]. Since the superconducting coherence length of Ta is relatively short (~100 nm), less than the channel length of the JJ, part of the JJ channel remained in the normal state, resulting in a finite on-resistance. In the new JJ device, Al, with a much longer superconducting coherence length (~1 µm), is used for the superconducting electrodes. To further enhance the interface transparency between the superconductor and semiconductor, epitaxial growth is employed to deposit the Al thin layer[30]. The Josephson junction (~250 nm length) is then patterned using electron-beam lithography, as shown in Figs. 6(a) and 6(b). Figure 6(c) presents the junction resistance $R_{XX}$ as a function of temperature. The drop in resistance at approximately 1.4 K corresponds to the Al electrodes becoming superconducting. Around 1.2 K, the junction resistance reaches zero, indicating that the JJ has entered the superconducting state. Thus, the nominal 1 K operating temperature considered for JFS-CryoMem lies within the superconducting regime demonstrated by the optimized junction. Figure 6(d) shows the current-voltage ($Idc - Vdc$) characteristics of the JJ, from which a critical current of $I_C \approx 60\ \mu A$ is extracted. The product $eI_C R_N \sim$ 1 meV exceeds the superconducting gap of the junction (~0.4 meV), demonstrating that a highly transparent junction has been achieved. Here, $R_N \approx 17\ \Omega$ is the normal-state resistance. The observation of a true supercurrent state confirms that zero on-resistance is achievable in optimized device structures. Although the optimized junction demonstrates that zero on-resistance is experimentally achievable, the device was characterized as an ungated Josephson junction and therefore does not provide the gate-dependent electrical characteristics required for JJFET-based cell selection. Consequently, the array-level analysis in this work uses the experimentally characterized gated JJFET of Fig. 1, for which gate-dependent I–V characteristics are available.

## Outlook and discussion

In summary, we have proposed and evaluated the JFS-CryoMem architecture, which combines a voltage-controlled JJFET selector with a nonvolatile FeSQUID storage element. Using experimentally calibrated compact models, we demonstrated cell-level write and read operations and verified selective access in array configurations. The architecture provides separate read and write paths. The estimated write energy is 27 fJ per cell. We further analyzed how array scaling affects the readout voltage, sense margin, required bitline current, and read energy. For an iso-sense margin, we calculated the required bitline current and average read energy for different array sizes (from 2×2 to 16 × 16 array). These results establish JFS-CryoMem as a promising device-array approach for nonvolatile cryogenic memory and identify the key sensing and energy tradeoffs that govern its array scaling.

## Data availability

The data that support the plots within this paper and other findings of this study are available from the corresponding author upon reasonable request.

## References:

1. Alam, S., Hossain, M. S., Srinivasa, S. R. & Aziz, A. Cryogenic memory technologies. *Nat. Electron. 2023 63* **6**, 185–198 (2023).
2. Islam, M. M., Alam, S., Hossain, M. S., Roy, K. & Aziz, A. A review of cryogenic neuromorphic hardware. *J. Appl. Phys.* **133**, (2023).
3. Bao, Z. *et al.* A cryogenic on-chip microwave pulse generator for large-scale superconducting quantum computing. *Nat. Commun. 2024 151* **15**, 5958- (2024).
4. Acharya, R. *et al.* Multiplexed superconducting qubit control at millikelvin temperatures with a low-power cryo-CMOS multiplexer. *Nat. Electron.* **6**, 900–909 (2023).
5. Islam, M. M. *et al.* Analog-to-Digital Converter Based on Voltage-controlled Superconducting Device. (2025).
6. Brennan, J. C. *et al.* Classical interfaces for controlling cryogenic quantum computing technologies. *APL Quantum* **2**, (2025).
7. Islam, M. M., Alam, S., Udoy, M. R. I., Hossain, M. S. & Aziz, A. A Cryogenic Artificial Synapse based on Superconducting Memristor. *Proc. ACM Gt. Lakes Symp. VLSI, GLSVLSI* 143–148 (2023) doi:10.1145/3583781.3590203.
8. Ferraris, A., Cha, E., Mueller, P., Moselund, K. & Zota, C. B. Cryogenic quantum computer control signal generation using high-electron-mobility transistors. *Commun. Eng. 2024 31* **3**, 146- (2024).
9. Takeuchi, N., Yamae, T., Yamashita, T., Yamamoto, T. & Yoshikawa, N. Microwave-multiplexed qubit controller using adiabatic superconductor logic. *npj Quantum Inf. 2024 101* **10**, 53- (2024).
10. Pellerano, S. *et al.* Cryogenic CMOS for Qubit Control and Readout. *Proc. Cust. Integr. Circuits Conf.* **2022-April**, (2022).
11. Xue, X. *et al.* CMOS-based cryogenic control of silicon quantum circuits. *Nat. 2021 5937858* **593**, 205–210 (2021).
12. Jeong, J. *et al.* Cryogenic III-V and Nb electronics integrated on silicon for large-scale quantum computing platforms. *Nat. Commun. 2024 151* **15**, 10809- (2024).
13. Tolpygo, S. K. Superconductor digital electronics: Scalability and energy efficiency issues (Review Article). *Low Temp. Phys.* **42**, 361–379 (2016).
14. Mamaluy, D. *et al.* Predictive first-principles simulations for co-designing next-generation energy-efficient AI systems. (2026).
15. Kalashnikov, D. S. *et al.* Demonstration of a Josephson vortex-based memory cell with microwave energy-efficient readout. *Commun. Phys. 2024 71* **7**, 88- (2024).
16. Patterson, R. L. *et al.* Electronic components and systems for cryogenic space applications. *AIP Conf. Proc.* **613**, 1585–1590 (2002).
17. Clark, P. E. *et al.* Ultra Low Temperature Ultra Low Power Instrument Packages for Planetary Surfaces. *AIP Conf. Proc.* **1208**, 541–548 (2010).
18. Ezzadeen, M. *et al.* Implementation of binarized neural networks immune to device variation and voltage drop employing resistive random access memory bridges and capacitive neurons. *Commun. Eng. 2024 31* **3**, 80- (2024).
19. Islam, M. M. *et al.* Harnessing Ferro-Valleytricity in Penta-Layer Rhombohedral Graphene for Memory and Compute. *Appl. Phys. Rev.* **12**, (2024).
20. Udoy, M. R. I., Alam, S., Islam, M. M., Jaiswal, A. & Aziz, A. A Review of Digital Pixel Sensors. *IEEE Access* **13**, 8533–8551 (2025).
21. Damsteegt, R. A., Overwater, R. W. J., Babaie, M. & Sebastiano, F. A Benchmark of Cryo-CMOS Embedded SRAM/DRAMs in 40-nm CMOS. *IEEE J. Solid-State Circuits* **59**, 2042–2054 (2024).
22. Nguyen, M. H. *et al.* Cryogenic Memory Architecture Integrating Spin Hall Effect based Magnetic Memory and Superconductive Cryotron Devices. *Sci. Reports 2020 101* **10**, 248- (2020).
23. Medeiros, O. *et al.* A scalable superconducting nanowire memory array with row–column addressing. *Nat. Electron. 2026 91* **9**, 69–77 (2026).
24. Volk, J., Wynn, A., Golden, E., Sherwood, T. & Tzimpragos, G. Addressable superconductor integrated circuit memory from delay lines. *Sci. Reports 2023 131* **13**, 16639- (2023).
25. Alam, S. *et al.* Cryogenic Memory Array based on Ferroelectric SQUID and Heater Cryotron. *Device Res. Conf. - Conf. Dig. DRC* **2022-June**, (2022).
26. Noah, G. M. *et al.* CMOS on-chip thermometry at deep cryogenic temperatures. *Appl. Phys. Rev.* **11**, (2024).
27. Barua, B. P., Udoy, M. R. I. & Aziz, A. A Review of Multiscale Thermal Modeling in Heterogeneous 3D ICs. *IEEE Access* (2026) doi:10.1109/ACCESS.2026.3697324.
28. Udoy, M. R. I. *et al.* ThermoPix: A High-Spatial-Resolution ElectronicPhotonic Temperature Sensor Array With Microsecond Row Readout. (2026).
29. Pan, W. *et al.* Quantum enhanced Josephson junction field-effect transistors for logic applications. *Mater. Sci. Eng. B* **310**, 117729 (2024).
30. Pan, W. *et al.* Epitaxial aluminum layer on antimonide heterostructures for exploring Josephson junction effects. *Mater. Sci. Eng. B* **318**, 118285 (2025).
31. Islam, M. M. *et al.* Reimagining Voltage-Controlled Cryogenic Boolean Logic Paradigm with Quantum-Enhanced Josephson Junction FETs. (2025).
32. Alam, S. *et al.* Cryogenic In-Memory Matrix-Vector Multiplication using Ferroelectric Superconducting Quantum Interference Device (FE-SQUID). *Proc. - Des. Autom. Conf.* **2023-July**, (2023).
33. Suleiman, M., Sarott, M. F., Trassin, M., Badarne, M. & Ivry, Y. Nonvolatile voltage-tunable ferroelectric-superconducting quantum interference memory devices. *Appl. Phys. Lett.* **119**, (2021).
34. Meyer, V., Sallese, J. M., Fazan, P., Bard, D. & Pecheux, F. Modeling the polarization in ferroelectric materials: a novel analytical approach. *Solid. State. Electron.* **47**, 1479–1486 (2003).
35. Alam, S., Islam, M. M., Hossain, M. S. & Aziz, A. Superconducting Josephson Junction FET-based Cryogenic Voltage Sense Amplifier. *Device Res. Conf. - Conf. Dig. DRC* **2022-June**, (2022).

## Acknowledgement

We thank Landon Schnebly for help in taking the SEM image. The work at Sandia National Laboratories (SNL) is supported by the LDRD program. SNL is a multi-mission laboratory managed and operated by the National Technology & Engineering Solutions of Sandia, LLC (NTESS), a wholly owned subsidiary of Honeywell International Inc., for the U.S. Department of Energy's National Nuclear Security Administration (DOE/NNSA) under Contract No. DE-NA0003525. This written work is authored by an employee of NTESS. The employee, not NTESS, owns the right, title, and interest in and to the written work and is responsible for its contents. Any subjective views or opinions that might be expressed in the written work do not necessarily represent the views of the U.S. Government. The publisher acknowledges that the U.S. Government retains a non-exclusive, paid-up, irrevocable, worldwide license to publish or reproduce the published form of this written work or allow others to do so for U.S. Government purposes. The DOE will provide public access to results of federally sponsored research in accordance with the DOE Public Access Plan.